\documentclass[twocolumn,aps,prd,preprintnumbers,showpacs,superscriptaddress,nofootinbib,amsmath,amssymb,floats,floatfix,showkeys,notitlepage,longbibliography]{revtex4-2}
\usepackage{orcidlink}
\usepackage{comment}
\usepackage{lipsum}
\usepackage{graphicx}
\usepackage{subfigure}
\usepackage{palatino}
\usepackage{sans}
\usepackage{hyperref}
\hypersetup{colorlinks=true,linkcolor=blue,urlcolor=blue,citecolor=blue}
\usepackage[toc,page]{appendix}
\usepackage[normalem]{ulem}
\usepackage{adjustbox}
\usepackage{latexsym}
\usepackage{amsmath}
\usepackage{amssymb}
\usepackage{amsfonts}
\usepackage{dcolumn}
\usepackage{bm}
\usepackage{tikz}
\usepackage{bigints}
\usepackage{array,tabularx, multirow,booktabs}
\usepackage[tracking=true]{microtype}
\SetTracking{}{500}
\SetTracking{encoding={*}, shape=sc}{40}
\UseRawInputEncoding 
\allowdisplaybreaks

\begin{document}
	\newcommand \nn{\nonumber}
	\newcommand \fc{\frac}
	\newcommand \lt{\left}
	\newcommand \rt{\right}
	\newcommand \pd{\partial}
	\newcommand \e{\text{e}}
	\newcommand \hmn{h_{\mu\nu}}
	\newcommand{\PR}[1]{\ensuremath{\left[#1\right]}} 
	\newcommand{\PC}[1]{\ensuremath{\left(#1\right)}} 
	\newcommand{\PX}[1]{\ensuremath{\left\lbrace#1\right\rbrace}} 
	\newcommand{\BR}[1]{\ensuremath{\left\langle#1\right\vert}} 
	\newcommand{\KT}[1]{\ensuremath{\left\vert#1\right\rangle}} 
	\newcommand{\MD}[1]{\ensuremath{\left\vert#1\right\vert}} 

\title{Observational signatures of quantum-corrected RN blackhole}

\author{Nikko John Leo S. Lobos \orcidlink{0000-0001-6976-8462}}
\email{nslobos@ust.edu.ph}
\affiliation{Electronics Engineering Department, University of Santo Tomas, Espa\~na Boulevard, Sampaloc, Manila 1008, Philippines}

\author{Virginia C. Fernandez}
\email{vcfernandez@tua.edu.ph}
\affiliation{Trinity University of Asia High School, Trinity University of Asia, Cathedral Heights, 275 E Rodriguez Sr. Ave, Quezon City, Metro Manila, Philippines}

\begin{abstract}
We investigate the observational signatures of a quantum-corrected Reissner-Nordstr\"om (RN) black hole to constrain Planck-scale modifications to spacetime geometry using current astrophysical data. By analyzing the null geodesic structure, we demonstrate that the quantum correction parameter, $\mathrm{a}$, acts as a repulsive geometric potential that opposes the gravitational compactification induced by the electric charge, $Q$. This competition leads to a parameter degeneracy wherein a highly charged, quantum-corrected black hole can mimic the shadow size of a classical Schwarzschild black hole. To resolve this, we employ the strong-field limit formalism to derive the deflection angle and the observables associated with relativistic Einstein rings. Our analysis reveals that while the electric charge enhances the deflection angle, the quantum correction suppresses it, providing a theoretical mechanism to distinguish the two effects. Confronting these predictions with the latest Event Horizon Telescope (EHT) observations, we derive robust constraints on the dimensionless parameter $\Pi = \mathrm{a}/Q$. We find that consistency with the shadow angular diameter of Sgr. A$^*$ requires $0 \le \Pi \lesssim 0.7$, implying that quantum geometric corrections cannot exceed approximately $70\%$ of the black hole charge without violating empirical bounds. These results highlight the potential of strong-field lensing to place precise phenomenological limits on quantum gravity candidates.
\end{abstract}	



\maketitle

\section{Introduction}\label{intro}
General Relativity (GR) has been remarkably successful in describing the gravitational dynamics of the universe, from the orbital precession of Mercury to the detection of gravitational waves. However, the theory predicts its own breakdown at space-time singularities, where curvature invariants diverge, and predictability is lost \cite{Hawking:1970zqf}. It is widely believed that these singularities are artifacts of the classical theory and may be resolved by a consistent theory of Quantum Gravity (QG).

In the absence of a complete unified theory, phenomenological models of regular or quantum-corrected black holes serve as a crucial testing ground. These effective field theories offer insights into how deviations from classical GR might manifest if the underlying geometry is modified by non-perturbative effects \cite{Ashtekar:2005qt, frisoni2023introduction}. Recent theoretical developments, including Loop Quantum Gravity (LQG), string theory, and asymptotically safe gravity, suggest that such corrections could alter the horizon structure and thermodynamic properties of black holes \cite{Maldacena:1997re, Reuter:2012id}. While perturbative quantum corrections are expected to be suppressed by the Planck scale, various non-perturbative models such as black hole mimickers, gravastars, or macroscopic quantum hairs suggest that geometric deformations could persist at horizon scales \cite{Nicolini:2008aj, Ali:2009zq, Konoplya:2019xmn}.

The advent of the Event Horizon Telescope (EHT) has opened a new window for testing these predictions \cite{EventHorizonTelescope:2019dse, EventHorizonTelescope:2019pgp, EventHorizonTelescope:2019ths}. The breakthrough imaging of the supermassive black holes M87$^*$ and Sgr. A$^*$ allows us to probe gravity in the strong-field regime with unprecedented precision. Consequently, the black hole shadow \cite{Perlick:2021aok} and strong-field gravitational lensing \cite{bozza2002gravitational, Tsukamoto:2016jzh, Banerjee:2019nnj} have emerged as powerful diagnostic tools. These observables can constrain generic deviations from the classical Kerr and Schwarzschild solutions, thereby placing bounds on modified gravity theories and bumpy black hole metrics \cite{Vagnozzi:2022moj, Bambi:2019tjh, Afrin:2022ztr, Khodadi:2024ubi}.

In this context, the Reissner-Nordstr\"om (RN) black hole provides a valuable testbed. Although astrophysical black holes are expected to be electrically neutral, charged solutions often serve as useful proxies for black holes with non-zero tidal charges, as in brane-world scenarios or for the effects of magnetic fields \cite{Wang:2021vbn, Ling:2021vgk, Zakharov:2014lqa, Pantig:2025eda, Ovgun:2025ctx}. Furthermore, the EHT collaboration has placed explicit constraints on charge-like parameters, motivating the continued study of RN-like geometries \cite{EventHorizonTelescope:2021dqv}.

In this work, we investigate the observational signatures of a quantum-corrected RN black hole, based on the metric introduced by Wu and Liu \cite{Wu:2022lqr}. While the correction parameter $\mathrm{a}$ in this metric is originally derived from the Planck length, we adopt a phenomenological approach, treating $\mathrm{a}$ as a free parameter characterizing the scale of geometric deformation. This allows us to test the metric as a prototype for regular black holes where repulsive corrections compete with the gravitational compactification induced by charge.

To address the hierarchy of scales, we introduce a dimensionless parameterization that explores the relative dominance of classical charge versus geometric corrections. This approach allows us to systematically analyze how repulsive deformations might manifest in the shadow radius and deflection angle, offering a standard for differentiating between classical charge effects and effective quantum inspired modifications. This paper builds upon recent advances in strong-field lensing formalism, utilizing Bozza's methodology to characterize deviations from classical predictions \cite{Bozza:2002zj}. Our goal is to determine whether the interplay between the electric charge $Q$ and the deformation parameter $\mathrm{a}$ produces distinguishable signatures that could be discerned by next-generation observational campaigns.

The paper is organized as follows: In Section \ref{sec:2}, we revisit the event horizon and photon sphere of the quantum-corrected black hole, utilizing our dimensionless parameter to examine the interplay of charge and corrections. In Section \ref{sec3}, we investigate the implications of the modified metric on the black hole shadow and compare our results with EHT bounds. In Section \ref{sec4}, we calculate the strong deflection angle of photons. Finally, in Section \ref{conc}, we summarize our findings.

\section{Quantum Corrected Blackhole Properties}\label{sec:2}
In this section, we derive the fundamental properties of the quantum-corrected Reissner-Nordstr\"om (RN) black hole and investigate the interplay between the electric charge $Q$ and the quantum correction parameter $\mathrm{a}$. The spacetime metric is given by \cite{Wu:2022lqr}:
\begin{equation}\label{eq.1}
    ds^{2} = -A(r)dt^{2}+B(r)dr^{2}+C(r)d\Omega^{2},
\end{equation} 
where the metric functions are defined as:
\begin{equation}\label{eq.2}
    A(r) = \frac{\sqrt{r^{2} - \mathrm{a}^{2}}}{r}-\frac{2M}{r}+\frac{Q^{2}}{r^2}, \quad B(r) = \frac{1}{A(r)}, \quad C(r) = r^2.
\end{equation} 
Here, $\mathrm{a} = 4\sqrt{\kappa} \equiv 4\ell_{p}$, where $\kappa$ is the gravitational coupling constant and $\ell_{p}$ is the Planck length. The parameter $\mathrm{a}$ represents the quantum excitation in the spherically symmetric spacetime \cite{Wu:2022lqr}.While theoretically $\mathrm{a}$ relates to the Planck length, for the purpose of astrophysical observation, we treat $\mathrm{a}$ as a free phenomenological parameter characterizing the scale of geometric deformation.

\subsection{Dimensionless Parameterization}
A major critique of phenomenological studies involving Planck-scale corrections is the vanishingly small magnitude of parameters like $\mathrm{a}$ in astrophysical regimes. To address this and to better isolate the qualitative effects of the quantum correction relative to the classical charge, we introduce a dimensionless parameter $\Pi$, defined as the ratio of the quantum parameter to the black hole charge:
\begin{equation}
    \Pi(\mathrm{a}, Q) \equiv \frac{\mathrm{a}}{Q}.
\end{equation}
Expanding $A(r)$ for small $\mathrm{a}$ ($\mathrm{a} \ll r$) and expressing terms using $\Pi$, we obtain an asymptotically approximated metric function:
\begin{equation}\label{eq.3}
    A(r)_{\text{eff}} \approx 1-\frac{2M}{r}+\frac{Q^2}{r^2}\left(1-\frac{\Pi}{2}\right) + \mathcal{O}(\mathrm{a}^4).
\end{equation} 
This parameterization offers a robust framework for analysis: if $\Pi \ll 1$, classical electromagnetic effects dominate; however, as $\Pi \to 1$, the regime where charge and quantum scales are comparable, quantum corrections may significantly alter the effective potential. This suggests that for black holes with small effective charges or tidal charges in brane scenarios, quantum geometric deformations could mimic or mask the signature of a standard RN charge.

\subsection{Event Horizon and Photon Sphere}
The event horizon, defined by the condition $A(r_{H}) = 0$, is the null hypersurface from which no information can escape. Solving for the roots of Eq. (\ref{eq.2}) yields the horizon radii:
\begin{equation}\label{eq.4} 
r_{H_{\pm}} = M \pm \frac{\sqrt{4M^2 - 4Q^2 + 2Q^2\Pi}}{2}.
\end{equation}
In the limit $\Pi \to 0$, we recover the standard RN horizons. Notably, Eq. (\ref{eq.4}) demonstrates that while the classical charge term $-4Q^2$ tends to shrink the horizon (potentially exposing a naked singularity if $Q > M$), the quantum term $+2Q^2\Pi$ acts repulsively, effectively increasing the horizon radius. This quantum shielding effect opposes the formation of naked singularities.

The photon sphere radius $r_{\text{ph}}$, critical for shadow formation and strong lensing, is determined by the condition:
\begin{equation}\label{eq.7}
    C'(r_{\text{ph}})A(r_{\text{ph}}) - C(r_{\text{ph}})A'(r_{\text{ph}}) = 0.
\end{equation}
Substituting the metric functions into Eq. (\ref{eq.7}) yields:
\begin{equation}\label{eq.8}
    r_{\text{ph}} = \frac{3M}{2} + \frac{\sqrt{9M^2 - 8Q^2 + 4Q^2\Pi^2}}{2}.
\end{equation}
Here again, we observe competing behaviors: the electric charge $Q$ reduces the photon sphere radius (strengthening gravity's pull), whereas the quantum parameter $\mathrm{a}$ (via $\Pi$) expands it. This expansion can be understood physically as a stiffening of the spacetime geometry due to quantum excitations, pushing the unstable photon orbit to a larger radius. This interplay is crucial for the observables discussed in the following sections, as it implies that a specific combination of charge and quantum correction could produce a shadow size indistinguishable from a Schwarzschild black hole, creating a degeneracy that only precise strong-lensing measurements might break.
  
\section{Blackhole Shadow of Quantum Corrected BH} \label{sec3}
The black hole shadow is a gravitationally lensed feature that provides a direct probe of the spacetime geometry in the strong-field regime. In this section, we analyze the shadow cast by the quantum-corrected RN black hole, specifically examining how the dimensionless parameter $\Pi$ modifies the shadow radius relative to the classical Reissner-Nordstr\"om predictions.

For a static, spherically symmetric metric, the radius of the shadow $R_{\text{sh}}$ as seen by a distant observer is determined by the critical impact parameter of photons orbiting at the photon sphere $r_{\text{ph}}$. This is given by the relation \cite{Perlick:2021aok, Cunha:2018acu}:
\begin{equation}\label{eq.9}
    R_{\text{sh}} = \frac{r_{\text{ph}}}{\sqrt{A(r_{\text{ph}})}},
\end{equation}
where $r_{\text{ph}}$ is the photon sphere radius derived in Eq. (\ref{eq.8}). For an observer at a finite radial distance $r_O$ (relevant for future near-field observations), the angular radius of the shadow $\alpha_{\text{sh}}$ is given by:
\begin{equation}\label{eq.11}
    \sin^2\alpha_{\text{sh}} = \frac{R_{\text{sh}}^2}{C(r_O)} A(r_O).
\end{equation}
However, for astrophysical black holes like M87* and Sgr A*, the observer is effectively at infinity ($r_O \to \infty$). In this limit, the shadow radius $R_{\text{sh}}$ corresponds to the apparent size of the dark region on the observer's sky.

Substituting the metric function $A(r)$ from Eq. (\ref{eq.2}) and the photon sphere radius $r_{\text{ph}}$ into Eq. (\ref{eq.9}), and expanding for small quantum corrections ($\Pi \ll 1$) and small charge, we obtain an analytical approximation for the shadow radius:
\begin{equation}\label{eq.12}
    R_{\text{sh}} \approx 3\sqrt{3}M \left( 1 - \frac{Q^2}{18M^2} + \frac{Q^2 \Pi}{36M^2} \right).
\end{equation}
Equation (\ref{eq.12}) encapsulates the core phenomenology of the model:
\begin{enumerate}
    \item Classical Limit ($Q \to 0$): The shadow radius reduces to the Schwarzschild value $R_{\text{sch}} = 3\sqrt{3}M$, as expected.
    \item Classical Charge Effect ($\Pi \to 0$): The term $-\frac{Q^2}{18M^2}$ represents the standard RN effect, where the electric charge strengthens the gravitational potential, pulling photon orbits inward and shrinking the shadow.
    \item Quantum Correction Effect ($\Pi > 0$): The term $+\frac{Q^2 \Pi}{36M^2}$ acts with the opposite sign. The quantum parameter $\mathrm{a}$ (and thus $\Pi$) effectively reduces the "compactness" of the horizon structure, acting as a repulsive geometric correction that enlarges the shadow.
\end{enumerate}

\subsection{Observational Constraints from EHT}
We now compare these theoretical predictions with the constraints provided by the Event Horizon Telescope (EHT). According to recent reports \cite{EventHorizonTelescope:2019dse, EventHorizonTelescope:2021dqv}, the bounds on the shadow radius are often expressed in terms of the fractional deviation $\delta$ from the Schwarzschild prediction:
\begin{equation}
    \delta = \frac{R_{\text{sh}}}{R_{\text{sch}}} - 1.
\end{equation}
For M87*, the EHT collaboration constrains the shadow size to be consistent with General Relativity within $\sim 17\%$, while for Sgr A*, the bounds are tighter, constraining deviations to within $\sim 10\%$ \cite{EventHorizonTelescope:2022wkp, Vagnozzi:2022moj}.

\begin{figure}[t]
    \centering
    \includegraphics[width=0.45\textwidth]{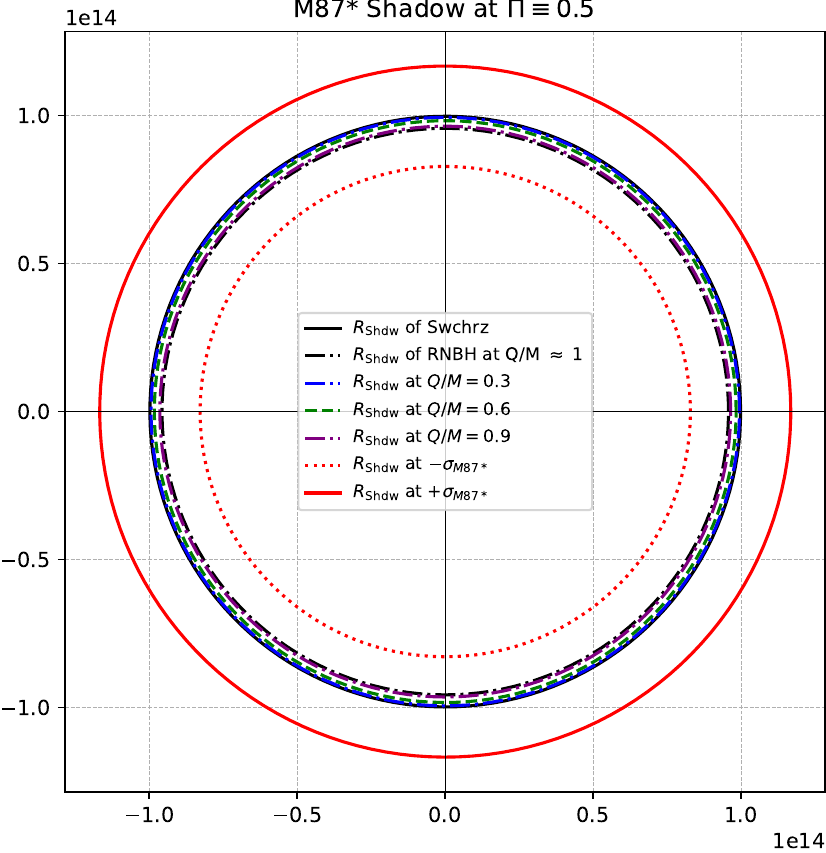}
    \includegraphics[width=0.45\textwidth]{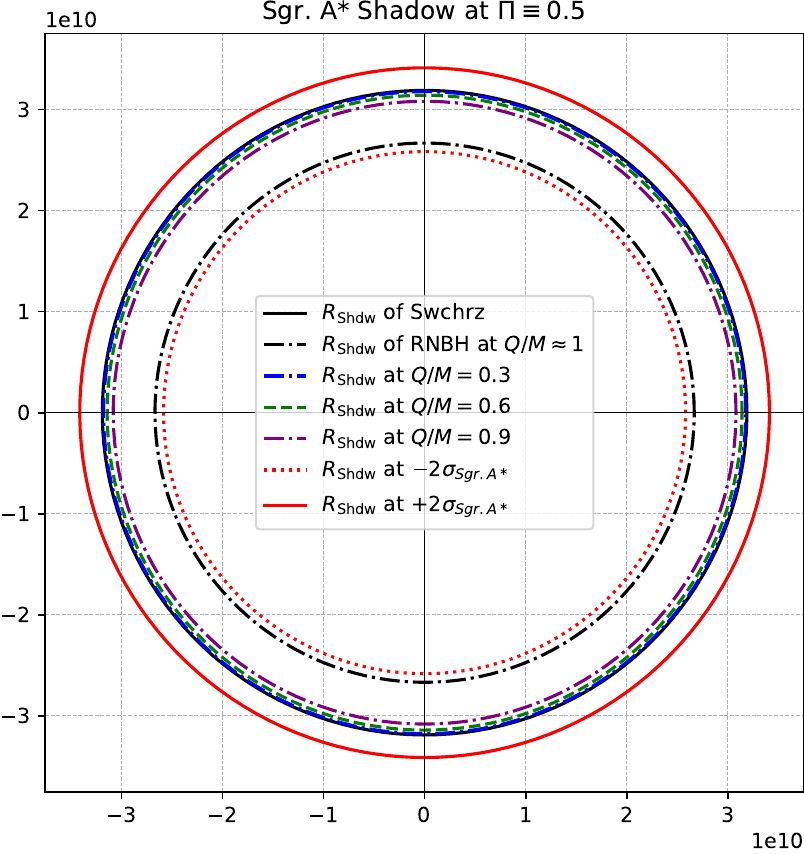}
    \caption{The variation of the shadow radius with electric charge $Q$ for the M87* black hole (above) and Sgr. A* (below). The solid red line corresponds to the EHT observational shadow radius upper bound, while the dashed lines indicate EHT observational lower bounds.}
    \label{fig:shadows}
\end{figure}

\begin{figure}[t]
    \centering
    \includegraphics[width=0.45\textwidth]{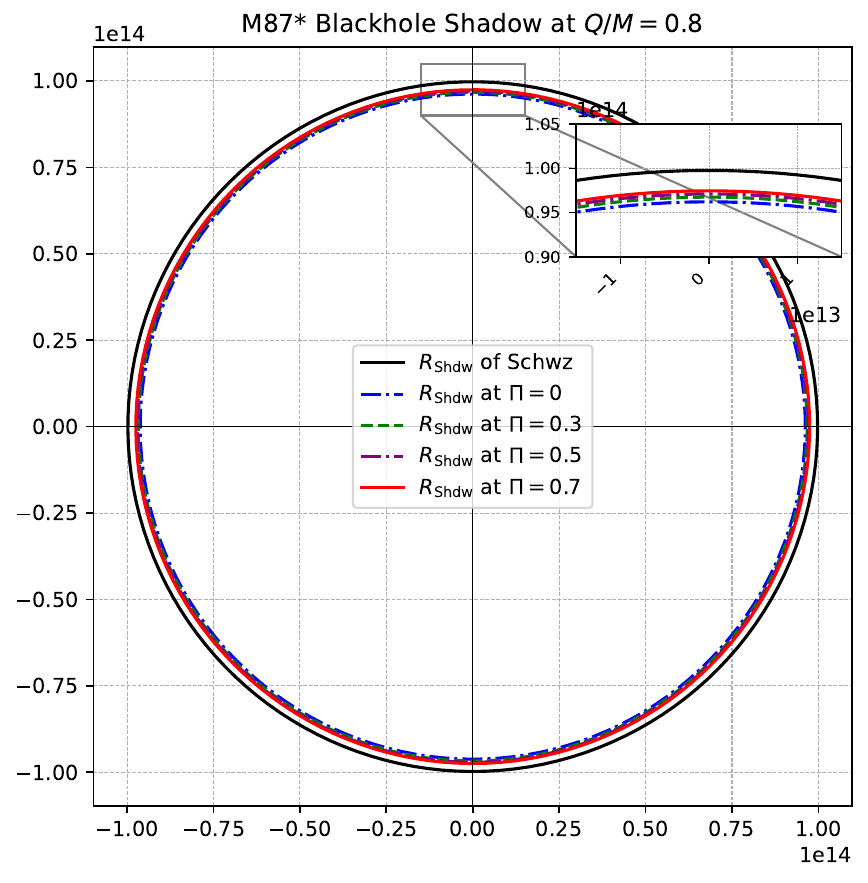}
    \includegraphics[width=0.45\textwidth]{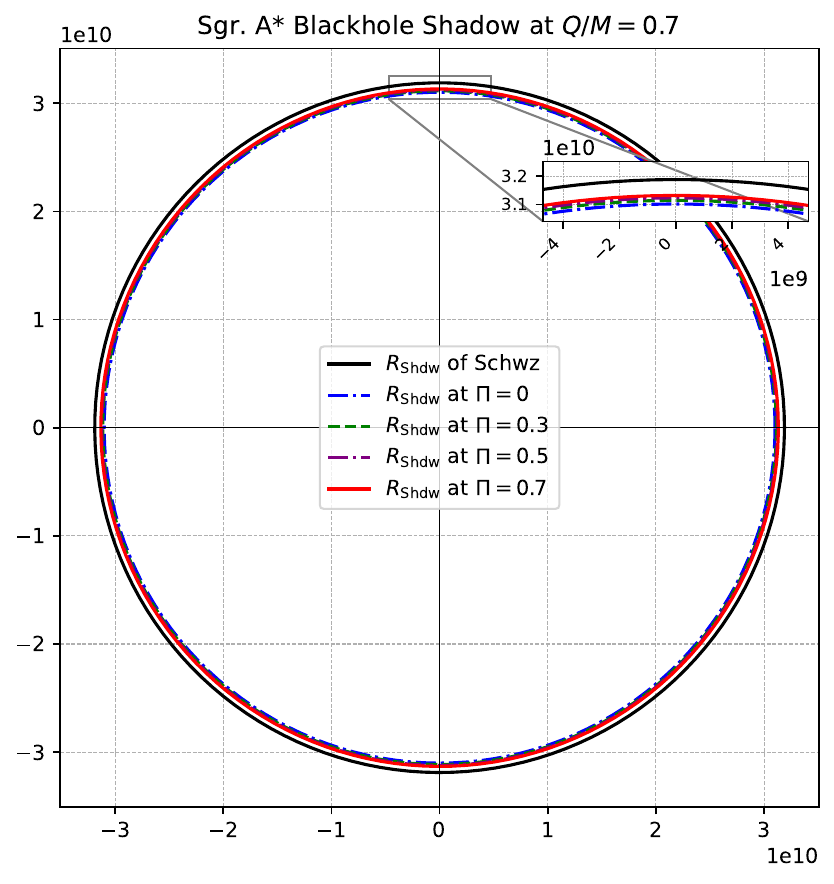}
    \caption{The variation of the shadow radius with quantum parameter $\Pi$ for the M87* black hole (above) and Sgr. A* (below). The dashed lines indicate EHT observational bounds.}
    \label{fig:shadows2}
\end{figure}

As illustrated in Fig. \ref{fig:shadows}, even for extremal values of the charge ($Q \to M$), the predicted shadow radius for \textcolor{black}{both the RN-Blackhole and} quantum-corrected RN black hole remains well within the confidence intervals established by the EHT.
\begin{itemize}
    \item Charge Suppression: Increasing $Q$ reduces the shadow size. For classical RN black holes, this reduction is often too small to be ruled out by current EHT resolution, provided $Q$ is not strictly extremal.
    \item Quantum Expansion: The introduction of the parameter $\Pi$ shifts the shadow radius back toward the Schwarzschild value. This implies that a highly charged black hole with significant quantum corrections could mimic a neutral Schwarzschild black hole, as shown in Fig. \ref{fig:shadows2}.
\end{itemize}

This degeneracy poses a challenge for observation: the fingerprint of the charge, Q, a smaller shadow, can be masked by the fingerprint of the quantum correction, a larger shadow. Consequently, while current EHT data is consistent with this metric, breaking the degeneracy between $Q$ and $\Pi$ requires precision measurements of the shadow shape or complementary data from strong gravitational lensing, which we explore in the next section.

\section{Analysis of the deflection angle in the strong field limit} \label{sec4}

This section investigates the strong gravitational lensing of null geodesics (photons) around a quantum-corrected Reissner-Nordstr\"om (RN) black hole. To quantify the deflection in the strong-field regime, we employ the formalism developed by Bozza \cite{bozza2002gravitational} with the modifications introduced by Tsukamoto \cite{Tsukamoto:2016jzh}. As described in the orbit equation solutions in Ref. \cite{tsukamoto2020deflection}, the total deflection angle $\alpha(r_0)$ as a function of the distance of closest approach $r_0$ is expressed as:
\begin{equation}
\label{eq.45}
\begin{split}
\alpha(r_{0}) &= I(r_{0}) - \pi \\
&= 2 \int^{\infty}_{r_{0}} \frac{dr}{\sqrt{\frac{R(r)C(r)}{B(r)}}}-\pi,
\end{split}
\end{equation}
where the function $R(r)$ is defined as:
\begin{equation}
\label{eq.44}
R(r) = \frac{A({r_0})r^2}{A(r)r_{0}^2}-1.
\end{equation}
Here, $A(r)$ corresponds to the metric function defined in Eq. (\ref{eq.2}), and $A(r_{0})$ represents the metric evaluated at the turning point $r_{0}$. To solve the integral in Eq. \eqref{eq.45}, we perform a series expansion around $r=r_{0}$ following the procedure in Ref. \cite{Tsukamoto:2016jzh}. The integral is decomposed into a regular part $I_{R}$ and a divergent part $I_{D}$. By introducing the variable $z \equiv 1-r_{0}/r$, the total integral $I(r_{0})$ yields:
\begin{equation}
I(r_{0}) = \int^{1}_{0}F(z, r_{0})dz = \int^{1}_{0}\left[F_{D}(z, r_{0})+F_{R}(z, r_{0})\right]dz,
\end{equation}
where $F(z, r_{0})$ represents the integrand separated into its divergent ($F_D$) and regular ($F_R$) components. The detailed expansion of Eq. \eqref{eq.45} is discussed in Refs. \cite{Tsukamoto:2016jzh, bozza2002gravitational}. Consequently, the strong deflection angle in the asymptotic limit is given by:
\begin{equation}
\label{eq.48}
\hat{\alpha}_{\text{str}} = -\bar{a} \log \left(\frac{b_0}{b_\text{crit}}-1\right)+\bar{b}+\mathcal{O}\left(\left(\frac{b_{0}}{b_{c}}-1\right)\log\left(\frac{b_{0}}{b_{c}}-1\right)\right),
\end{equation}
where $\bar{a}$ and $\bar{b}$ are the strong deflection coefficients, while $b_{0}$ and $b_\text{crit}$ denote the impact parameter at the closest approach $r_{0}$ and the critical impact parameter, respectively. The first term in Eq. \eqref{eq.48} arises from the diverging integral, dominating near the photon sphere, while the second term captures the contribution from the regular integral. The coefficients are explicitly defined as \cite{tsukamoto2020deflection}:
\begin{equation}
\label{eq.49}
\bar{a} = \sqrt{\frac{2}{2A(r_\text{ps}) - A''(r_\text{ps})r_\text{ps}^2}},
\end{equation}
and
\begin{equation}
\label{eq.50}
\bar{b} = \bar{a} \log\left[r_\text{ps} \left( \frac{2}{r_\text{ps}^2}-\frac{A''(r_\text{ps})}{A(r_\text{ps})}\right) \right]+I_{R}(r_\text{ps})-\pi,
\end{equation}
where the subscript ``ps'' denotes evaluation at the photon sphere radius $r_{\text{ps}}$ (or $r_{\text{ph}}$), and the double prime indicates the second derivative with respect to the radial coordinate.

Evaluating $\bar{a}$ and the logarithmic argument of $\bar{b}$ using the metric functions yields:
\begin{equation}
    \begin{split}\label{19}
        \bar{a} &= \sqrt{\frac{r^{2}_{\rm ph}}{r^{2}_{\rm ph}-2Q^2+Q^2\Pi}},\\
        \bar{b} & =\log\left[ \frac{(8 - 4\Pi)Q^2-4r^2_{\rm ph}}{4Mr_{\rm ph} +(\Pi-2)Q^2 -2r^2_{\rm ph}} \right]+I_{R}(r_{\rm ph})-\pi.
    \end{split}
\end{equation}
Notably, in the limit $Q \rightarrow 0$ and $\mathrm{a} \rightarrow 0$ (implying $\Pi \to 0$), these expressions recover the standard Schwarzschild strong deflection coefficients found in Ref. \cite{bozza2002gravitational}. The regular integral term $I_{R}$ is defined as:
\begin{equation}
\label{eq.53}
I_{R}(r_{0}) \equiv \int^{1}_{0}\left[\tau_{R}(z, r_{0}) - \tau_{D}(z,r_{0})\right] dz,
\end{equation}
where $\tau_{R}(z, r_{0})$ is derived from the trajectory expansion \cite{tsukamoto2020deflection}:
\begin{equation}
\label{eq.54}
\tau_{R}(z, r_{0}) =\frac{2r_{0}}{\sqrt{G(z, r_{0})}}.
\end{equation}
Here, $G(z, r_{0})$ encapsulates the metric functions $R, C, A$. Evaluating at the photon sphere $r_{0}=r_\text{ps}$, this becomes:
\begin{equation}
\label{eq.55}
\tau_{R}(z, r_\text{ps}) =\frac{2r_\text{ps}}{\sqrt{\sum_{m=2}^{k}c_{m}(r_\text{ps})z^{m}}}.
\end{equation}
The divergent term is given by:
\begin{equation}
\label{eq.56}
\tau_{D}(z, r_\text{ps}) = \frac{2r_\text{ps}}{\sqrt{c_{2}z^{2}}},
\end{equation}
where $c_m$ are the expansion coefficients. The evaluation of $I_{R}(r_{\text{ps}})$ yields:
\begin{equation}\label{24}
    I_{R}(r_{\text{ps}}) = \frac{2}{\sqrt{c_{3}}}\ln\left(\frac{4c_{3}}{2\sqrt{c_3 (c_1+c_2+c_3)}+c_2 +2c_3} \right),
\end{equation}
with the coefficients determined by the quantum-corrected metric parameters:
\begin{equation}\label{25}
    \begin{split}
        c_1 &= 4Q^{2}(\Pi-2),\\
        c_2 &=-8\left[2Q^{2}(\Pi-2) + \left(3M+\sqrt{(4\Pi-8)Q^{2}+9M^{2}}\right)M\right],\\
        c_3 &= -8\left[-2Q^{2}(\Pi-2) -\frac{3M(3M+\sqrt{(4\Pi-8)Q^{2}+9M^{2}})}{2}\right].
    \end{split}
\end{equation}
Finally, the impact parameter at closest approach is given by \cite{Pantig:2022gih}:
\begin{equation}
    b_{0}^{2} = \frac{2r_{0}^{3}}{r_0-M}.
\end{equation}
As $r_{0}\rightarrow r_{\text{ph}}$, $b_0$ approaches the critical impact parameter $b_{\text{crit}}$.

\begin{figure}[t]
    \centering
    \includegraphics[width=0.48\textwidth]{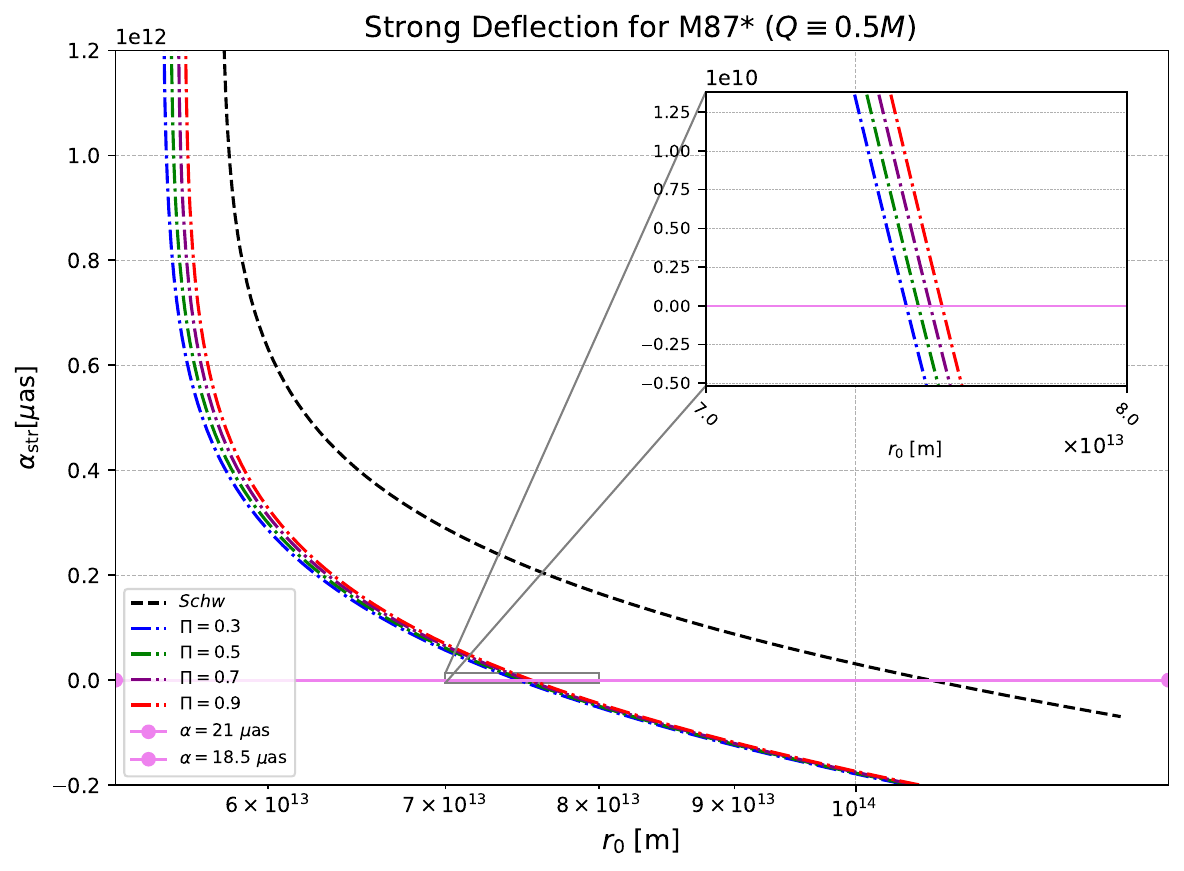}
    \includegraphics[width=0.48\textwidth]{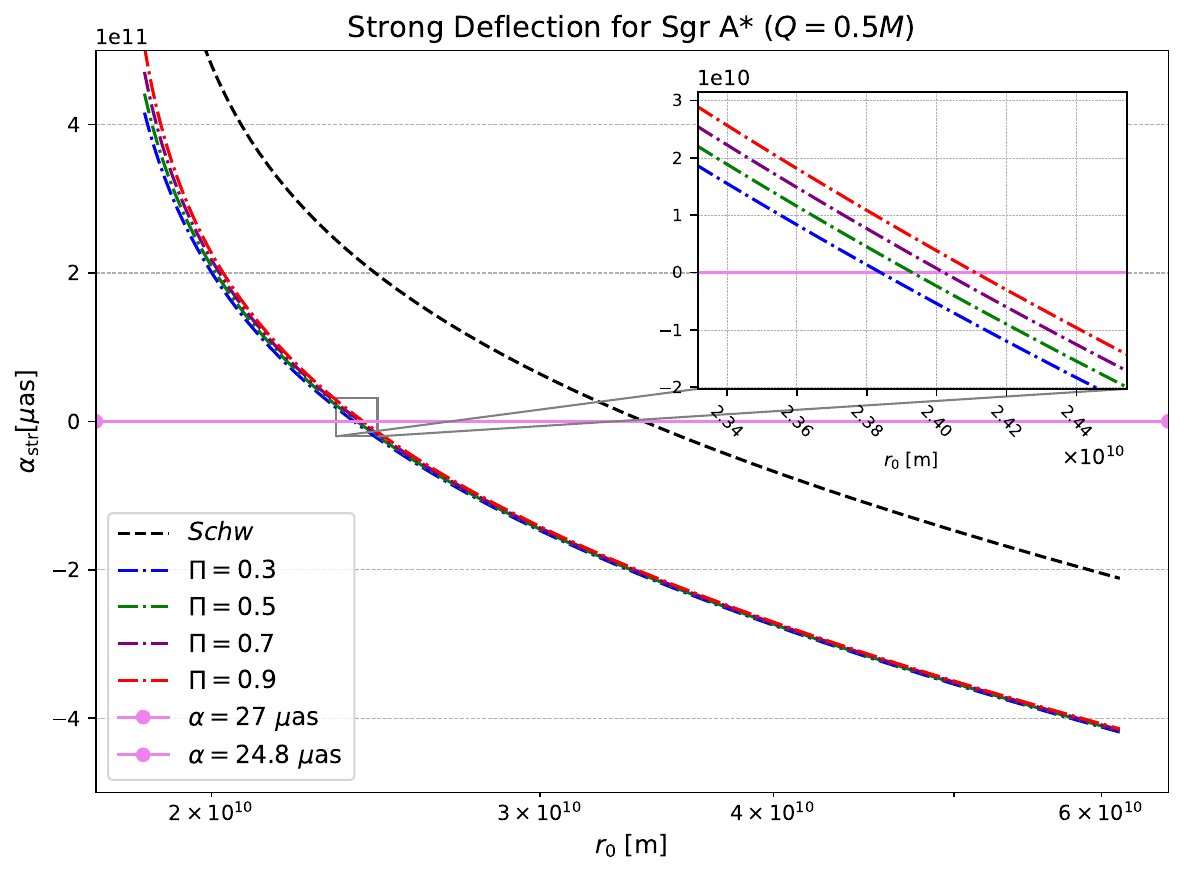}
    \caption{The behavior of the strong deflection angle $\alpha(r_0)$ as a function of the closest approach distance $r_0$ for M87* (top) and Sgr A* (bottom). The vertical asymptotes indicate the location of the photon sphere for varying values of $\Pi$. The shift in these asymptotes corresponds to the variation in shadow size constrained by EHT observations.}
\label{fig:3}

\end{figure}

The numerical results of the strong deflection analysis are presented in Fig. \ref{fig:3}. The plots illustrate the deflection angle $\alpha$ as a function of the distance of closest approach $r_0$ for M87* (top) and Sgr A* (bottom). The vertical asymptotes in the figure correspond to the photon sphere radius where the deflection angle diverges logarithmically.

We observe that the introduction of the quantum parameter $\Pi$ shifts the position of this divergence (and thus the critical impact parameter $b_{\text{crit}}$) outward compared to the Schwarzschild case (black dashed curve). This confirms our earlier analysis that the quantum correction acts as a repulsive geometric potential. The observational constraints from the EHT, which limit the angular shadow size $\theta_{\text{sh}}$, directly constrain the location of these vertical asymptotes. Specifically, for a model to be consistent with observations, the critical impact parameter derived from the strong deflection limit must correspond to a shadow angular radius within the ranges \( 18.5\ \mu\text{as} < \theta_{\text{M87*}} < 21\ \mu\text{as} \) \cite{EventHorizonTelescope:2019dse} and \( 24.8\ \mu\text{as} < \theta_{\text{Sgr A*}} < 27\ \mu\text{as} \) \cite{EventHorizonTelescope:2022wkp}.

Although the deviations introduced by $\Pi$ are subtle at the current observational resolution, the monotonic shift in the divergence point suggests that the quantum correction leaves a distinct imprint on the asymptotic structure of the spacetime. While current EHT data is consistent with the classical limit, the presence of $\Pi$ modifies the specific photon ring geometry in a way that could be distinguishable with the higher sensitivity expected from future next-generation VLBI missions. Thus, our analysis highlights that even Planck-scale corrections can theoretically manifest in macroscopic lensing phenomena, providing a potential pathway for testing quantum gravity candidates in the strong-field regime.

\section{Conclusions} \label{conc}
In this work, we have performed a comprehensive investigation of the optical signatures of a quantum-corrected Reissner-Nordstr\"om (RN) black hole, analyzing the interplay between the electric charge $Q$ and the quantum correction parameter $\mathrm{a}$. Motivated by the search for observational footprints of Planck-scale physics in astrophysical regimes, we utilized the strong-field limit formalism and recent constraints from the Event Horizon Telescope (EHT) regarding M87$^*$ and Sgr.$^*$ to assess the phenomenological viability of this spacetime.

Our analysis of the black hole shadow revealed a distinct competitive mechanism between classical and quantum effects.  While the classical electric charge acts to deepen the gravitational potential thereby compactifying the photon orbits and shrinking the shadow the quantum parameter, $\mathrm{a}$, introduces a repulsive geometric correction. By defining a dimensionless parameter $\Pi = \mathrm{a}/Q$, we demonstrated that quantum corrections effectively shield the horizon, pushing the photon sphere outward. This behavior creates a parameter degeneracy; a highly charged black hole with significant quantum corrections can mimic the shadow size of a neutral Schwarzschild black hole, making shadow radius measurements alone insufficient to disentangle these effects within current observational uncertainties.

To break this degeneracy and constrain the model, we examined the strong gravitational lensing of null geodesics. Our numerical results indicate that while the electric charge enhances the deflection angle near the photon sphere, the quantum correction parameter $\mathrm{a}$ suppresses it. By mapping the theoretical shadow radius derived from the strong deflection limit onto the observational bounds of the photon ring, we established a concrete constraint on the theory.

Crucially, our analysis yields a consistent upper bound for the quantum correction parameter across both supermassive black hole candidates. Comparison with the EHT data for Sgr A*, which provides the tightest observational constraints, restricts the dimensionless parameter to the range:
\begin{equation}
0 \le \Pi \lesssim 0.7.
\end{equation}
This same bound is fully consistent with the M87$^*$ data, albeit with looser margins. The constraint $\Pi < 0.7$ applies to macroscopic deviations. This consistency implies that for a physically viable quantum-corrected RN model, the quantum geometric correction cannot exceed approximately $70\%$ of the black hole's charge. Values of $\Pi$ exceeding this threshold would produce a shadow diameter larger than the $27\, \mu\text{as}$ upper limit observed for Sgr.$^*$ and the corresponding bounds for M87*, regardless of the specific strong deflection asymptotic behavior.

\textcolor{black}{This study suggests several important areas for future research. The unique geometric stiffening of spacetime caused by the quantum correction requires a thorough investigation of time-dependent perturbative phenomena. This includes studying the quasinormal mode (QNM) spectrum and the dynamic stability of the event horizon. Most crucially, extending this theoretical framework to a more realistic astrophysical regime requires including angular momentum. Future studies should apply the Newman-Janis algorithm to construct rotating counterparts to this metric and incorporate broader classes of Generalized Uncertainty Principles (GUP). This extension will be vital for determining whether spin quantum coupling mechanisms natively enhance the observational detectability of Planck-scale physics. The introduction of the dimensionless spin parameter will invariably interact with the quantum parameter, generating a complex, multi-parameter observational degeneracy. In such a rotating framework, the breaking of spherical symmetry yields an oblate, D-shaped shadow. However, because frame-dragging causes a measurable left-right difference in both the main shadow boundary and the strong-deflection lensing observations, precise measurements of shadow oblateness and the asymmetric displacement of higher-order photon rings could theoretically provide the independent constraints needed to distinguish fundamental quantum geometric changes from standard astrophysical spin.}

In summary, we conclude that even minimal corrections stemming from the quantization of spacetime leave a distinct imprint on the asymptotic structure of the geometry. While current observations are consistent with the classical limit, the theoretical bounds derived here—specifically $\Pi \lesssim 0.7$—provide a clear target for next-generation high-resolution interferometry, offering a robust pathway to test quantum gravity candidates in the strong-field regime.

\section{Acknowledgements}
N.J.L. Lobos and V.C. Fernandez gratefully acknowledge the University of Santo Tomas and Trinity University of Asia for their continued support and encouragement of our research endeavors. We especially thank both institutions for providing an environment that fosters academic inquiry and for allowing us to pursue this study.
\bibliography{references}

\end{document}